# Lotka's Law and Pattern of Author Productivity in the Field of Brain Concussion Research: A Scientometric Analysis

S.Roselin Jahina
rosejsshaki@gmail.com

Dr. M.Sadik Batcha
*Annamalai University*

Muneer Ahmad
*Annamalai University*, muneerbangroo@gmail.com

# Lotka's Law and Pattern of Author Productivity in the Field of Brain Concussion Research: A Scientometric Analysis

S Roselin Jahina[1], Dr. M. Sadik Batcha[2], Muneer Ahmad[3]

**S.Roselin Jahina[1],** *Ph.D Research Scholar, Department of Library and Information Science, Annamalai University, Tamil Nadu, India – 608002, email- rosejsshaki@gmail.com*
**Dr. M.Sadik Batcha[2],** *Mentor, Professor & University Librarian, Department of Library and Information Science, Annamalai University, Tamil Nadu, India – 608002, email- msbau@rediffmail.com*
**Muneer Ahmad[3]**, *Ph.D Research Scholar, Department of Library and Information Science, Annamalai University, Tamil Nadu, India – 608002, email-muneerbangroo@gmail.com*

**Abstract**

The present study deals a scientometric analysis of 8486 bibliometric publications retrieved from the Web of Science database during the period 2008 to 2017. Data is collected and analyzed using Bibexcel software. The study focuses on various aspect of the quantitative research such as growth of papers (year wise), Collaborative Index (CI), Degree of Collaboration (DC), Co-authorship Index (CAI), Collaborative Co-efficient (CC), Modified Collaborative Co-Efficient (MCC), Lotka's Exponent value, Kolmogorov-Smirnov test (K-S Test).

**Keywords:** Scientometrics, Brain Concussion, Collaborative Index (CI), Degree of Collaboration (DC), Co-authorship Index (CAI), Collaborative Co-efficient (CC), Modified Collaborative Co-Efficient (MCC), Lotka's Exponent value, Kolmogorov-Smirnov test (K-S Test)

## 1. Introduction

Scientometrics defined as the "quantitative study of science, communication in science, and science policy" (Hess, 1997)[1]. Scientometrics developed at a distance from the sociology of science and closer to the library and the information sciences. At the same time, the value of scientometric indicators to inform scientific policies and the management of research has become evident (Irvine & Martin, 1984)[2]. A brain injury caused by a blow to the head or a violent shaking of the head and body. This occurs from a mild blow to the head, either with or without loss of consciousness, and can lead to temporary cognitive symptoms. Symptoms may include headache, confusion, lack of coordination, memory loss, nausea, vomiting, dizziness, ringing in the ears, sleepiness and excessive fatigue. There's no specific cure for concussion. Rest and restricting activities allow the brain to recover. This means that one should temporarily reduce time spent on sports, video games, TV or too much socializing. Medication for headache pain or ondansetron or other anti-nausea medication can be used for symptoms.

## 2. Review of Literature

There have been enormous amount of scientometric studies all across the world. Some of the relevant studies in the aforesaid direction are worthy of examinations. (Batcha & Ahmad, 2017)[3] analysed comparative analysis of Indian Journal of Information Sources and Services (IJISS) and Pakistan Journal of Library and Information Science (PJLIS) during 2011-2017 and studied various aspects like year wise distribution of papers, authorship pattern & author productivity, degree of collaboration pattern of Co-Authorship , average length of papers , average keywords, etc and  found 138 (94.52%) of contributions from IJISS were made by Indian authors and similarly 94 (77.05) of contributions from PJLIS were done by Pakistani authors. Papers by Indian and Pakistani Authors with Foreign Collaboration are minimal (1.37% of articles) and (4.10% of articles) respectively.

(Batcha, Jahina, & Ahmad, 2018)[4] has examined scientometric analysis of the DESIDOC Journal and analyzed the pattern of growth of the research output published in the journal, pattern of authorship, author productivity, and, subjects covered to the papers over the period (2013-2017). It found that 227 papers were published during the period of study (2001-2012). The maximum numbers of articles were collaborative in nature. The subject concentration of the journal noted was Scientometrics. The maximum numbers of articles (65 %) have ranged their thought contents between 6 and 10 pages.

(Ahmad & Batcha, 2019)[5] analyzed research productivity in Journal of Documentation (JDoc) for a period of 30 years between 1989 and 2018. Web of Science database a service from Clarivate Analytics has been used to download citation and source data. Bibexcel and Histcite application software have been used to present the datasets. Analysis part focuses on the parameters like citation impact at local and global level, influential authors and their total output, ranking of contributing institutions and countries. In addition to this scientographical mapping of data is presented through graphs using VOSviewer software mapping technique.

(Ahmad, Batcha, Wani, Khan, & Jahina, 2017)[6] explored scientometric analysis of the Webology Journal. The paper analyses the pattern of growth of the research output published in the journal, pattern of authorship, author productivity, and subjects covered to the papers over the period (2013-2017). It was found that 62 papers were published during the period of study (2013-2017).

The maximum numbers of articles were collaborative in nature. The subject concentration of the journal noted was Social Networking/Web 2.0/Library 2.0 and Scientometrics or Bibliometrics. Iranian researchers contributed the maximum number of articles (37.10%). The study applied standard formula and statistical tools to bring out the factual results.

(Ahmad & Batcha, 2019)[7] studied the scholarly communication of Bharathiar University which is one of the vibrant universities in Tamil Nadu. The study find out the impact of research produced, year-wise research output, citation impact at local and global level, prominent authors and their total output, top journals of publications, collaborating countries, most contributing departments and publication trends of the university during 2009 to 2018. The 10 years' publication data of the university indicate that a total of 3440 papers have been published from 2009 to 2018 receiving 38104 citations with h-index as 68. In addition the study used scientographical mapping of data and presented it through graphs using VOSviewer software mapping technique.

(Ahmad, Batcha, & Jahina, 2019)[8] quantitatively identified the research productivity in the area of artificial intelligence at global level over the study period of ten years (2008-2017). The study identified the trends and characteristics of growth and collaboration pattern of artificial intelligence research output. Average growth rate of artificial intelligence per year increases at the rate of 0.862. The multi-authorship pattern in the study is found high and the average number of authors per paper is 3.31. Collaborative Index is noted to be the highest range in the year 2014 with 3.50. Mean CI during the period of study is 3.24. This is also supported by the mean degree of collaboration at the percentage of 0.83 .The mean CC observed is 0.4635. Lotka's Law of authorship productivity is good for application in the field of artificial intelligence literature. The distribution frequency of the authorship follows the exact Lotka's Inverse Law with the exponent á = 2. The modified form of the inverse square law, i.e., Inverse Power Law with á and C parameters as 2.84 and 0.8083 for artificial intelligence literature is applicable and appears to provide a good fit. Relative Growth Rate [Rt(P)] of an article gradually increases from -0.0002 to 1.5405, correspondingly the value of doubling time of the articles Dt(P) decreases from 1.0998 to 0.4499 (2008-2017). At the outset the study reveals the fact that the artificial intelligence literature research study is one of the emerging and blooming fields in the domain of information sciences.

(Batcha, Dar, & Ahmad, 2019)[9] presented a scientometric analysis of the journal titled "Cognition" for a period of 20 years from 1999 to 2018. The study was conducted with an aim to provide a summary of research activity in the journal and characterize its most aspects. The research coverage includes the year wise distribution of articles, authors, institutions, countries and citation analysis of the journal. The analysis showed that 2870 papers were published in journal of Cognition from 1999 to 2018. The study identified top 20 prolific authors, institutions and countries of the journal. Researchers from USA have made the most percentage of contributions.

**3. Objective of the study**

- To quantify the research output in the form of publications and average growth rate of literature in the field of Brain Concussion over the study period of ten years (2008-2017).
- To analysis the authorship pattern and degree of collaboration of research in the field of Brain Concussion during the period of study.
- To analyze the research trend with collaborative co-efficient, Modulated Collaborative Co-efficient and Collaborative Index in the global literature of Brain Concussion.
- The study the growth trend with the investigation of Relative Growth Rate (RGR) of distributions.
- To discover the Doubling Time (DT) for the productions to turn out to be double of the current sum.
- To test the applicability of Lotka's Law in the author productivity.
- To analyze whether "n" worth affirms to Lotka's Law through K-S Test.

**4. Methodology**

The data presented in this paper have been accessed from Web of Science published by Clarivate Analytics. The basic data relating to total publications during 2008-2017, has been collected in the month of January 2018 using Web of Science database. The searches were performed on the name of Brain Concussion using Basic search term on Web of Science Core Collection with all probabilities and bibliographical details amounting of 8486 research papers collectively contributed by 41264 authors. All the searched results were saved in .txt files and then imported into Bibexcel and VOSviewer to organize, analyze and generate the tables, graphs and charts for final study.

## 5. Analysis and Interpretation of the Result

Table 1: Year wise Distribution and Average Growth Rate of Publications in Brain Concussion

| S.NO | Year | Res.Output | % | Cum.Output | Cum.% | Growth Rate |
|---|---|---|---|---|---|---|
| 1 | 2008 | 331 | 3.90 | 331 | 3.9 | - |
| 2 | 2009 | 477 | 5.62 | 808 | 9.52 | 0.694 |
| 3 | 2010 | 487 | 5.74 | 1295 | 15.26 | 0.979 |
| 4 | 2011 | 583 | 6.87 | 1878 | 22.13 | 0.835 |
| 5 | 2012 | 769 | 9.06 | 2647 | 31.19 | 0.758 |
| 6 | 2013 | 862 | 10.16 | 3509 | 41.35 | 0.892 |
| 7 | 2014 | 1026 | 12.09 | 4535 | 53.44 | 0.840 |
| 8 | 2015 | 1125 | 13.26 | 5660 | 66.7 | 0.912 |
| 9 | 2016 | 1332 | 15.70 | 6992 | 82.4 | 0.845 |
| 10 | 2017 | 1494 | 17.61 | 8486 | 100 | 0.892 |
|  | **Total** | **8486** | **100%** |  |  | **0.850** |

Table 1 describes the growth of research publications published in the field of Brain Concussion during the study period of 2008-2017. Totally 8486 publications were published. The highest number of articles, 1494 (17.61%) were published in the year 2017. The second highest numbers of articles were published in the year 2016 (15.70%).

Table 2: Analysis of Authorship Pattern among the scientists of Brain Concussion

| Authors | 2008 | 2009 | 2010 | 2011 | 2012 | 2013 | 2014 | 2015 | 2016 | 2017 | Total | % | Total Authors |
|---|---|---|---|---|---|---|---|---|---|---|---|---|---|
| 1 | 29 | 33 | 38 | 49 | 37 | 56 | 41 | 55 | 56 | 7 | 401 | 5.36 | 401 |
| 2 | 45 | 72 | 65 | 75 | 81 | 89 | 107 | 125 | 153 | 32 | 844 | 11.28 | 1688 |
| 3 | 62 | 90 | 63 | 90 | 121 | 114 | 148 | 145 | 157 | 46 | 1036 | 13.85 | 3108 |
| 4 | 51 | 72 | 76 | 87 | 120 | 125 | 147 | 144 | 199 | 51 | 1072 | 14.33 | 4288 |
| 5 | 38 | 74 | 69 | 61 | 119 | 130 | 156 | 165 | 171 | 61 | 1044 | 13.95 | 5220 |
| 6 | 39 | 46 | 65 | 71 | 92 | 105 | 120 | 136 | 158 | 60 | 892 | 11.92 | 5352 |
| 7 | 19 | 41 | 34 | 43 | 60 | 70 | 79 | 95 | 127 | 57 | 625 | 8.35 | 4375 |
| 8 | 18 | 23 | 35 | 39 | 47 | 52 | 61 | 83 | 64 | 37 | 459 | 6.13 | 3672 |
| 9 | 11 | 8 | 16 | 29 | 30 | 32 | 59 | 49 | 68 | 42 | 344 | 4.60 | 3096 |
| 10 | 6 | 10 | 11 | 12 | 26 | 28 | 29 | 44 | 47 | 22 | 235 | 3.14 | 2350 |
| 11 | 6 | 3 | 5 | 10 | 11 | 16 | 28 | 25 | 35 | 14 | 153 | 2.04 | 1683 |
| 12 | 4 | - | 1 | 4 | 6 | 9 | 14 | 13 | 27 | 16 | 94 | 1.26 | 1128 |
| 13 | 1 | 1 | 5 | 7 | 2 | 12 | 9 | 9 | 24 | 14 | 84 | 1.12 | 1092 |
| 14 | - | - | - | 3 | 4 | 3 | 8 | 7 | 12 | 7 | 44 | 0.59 | 616 |
| 15 | - | 1 | 2 | - | 2 | 4 | 2 | 2 | 7 | 6 | 26 | 0.35 | 390 |
| 16 | 1 | 1 | 2 | - | 4 | 2 | 2 | 5 | 2 | 1 | 20 | 0.27 | 320 |
| 17 | - | - | - | - | - | 3 | 3 | 2 | 8 | 3 | 19 | 0.25 | 323 |
| 18 | - | 1 | - | - | - | - | - | 3 | 2 | 3 | 9 | 0.12 | 162 |
| 19 | - | - | - | - | 3 | 1 | 1 | 3 | 4 | - | 12 | 0.16 | 228 |
| 20 | - | - | - | 1 | - | 1 | 5 | 3 | 3 | 3 | 16 | 0.21 | 320 |
| 21 | - | - | - | - | - | - | 2 | 2 | 2 | 2 | 8 | 0.11 | 168 |
| - | - | - | - | - | - | 1 | - | 2 | - | - | 1 | 4 | 0.05 | 88 |
| 23 | - | - | - | - | - | 3 | - | 2 | - | - | 5 | 0.07 | 115 |
| 24 | - | - | - | - | - | 1 | - | - | 1 | 1 | 3 | 0.04 | 72 |
| 25 | - | - | - | - | - | - | 2 | - | - | 3 | 5 | 0.07 | 125 |
| 26 | - | - | - | - | - | - | - | 2 | 1 | 1 | 4 | 0.05 | 104 |
| 27 | - | - | - | - | 1 | - | - | - | - | 2 | 3 | 0.04 | 81 |
| 28 | - | - | - | - | - | 6 | - | 1 | - | - | 7 | 0.09 | 196 |
| 29 | - | - | - | - | - | - | - | - | 1 | - | 1 | 0.01 | 29 |
| 30 | - | 1 | - | - | - | - | - | - | - | - | 1 | 0.01 | 30 |
| 31 | - | - | - | - | - | - | - | - | - | 1 | 1 | 0.01 | 31 |
| 32 | - | - | - | - | - | - | - | - | 1 | - | 1 | 0.01 | 32 |
| 34 | - | - | - | - | 1 | - | - | - | 1 | - | 2 | 0.03 | 68 |
| 35 | - | - | - | - | - | - | - | 3 | - | - | 3 | 0.04 | 105 |
| 36 | - | - | - | - | - | - | - | - | - | 1 | 1 | 0.01 | 36 |
| 37 | - | - | - | - | - | - | - | 1 | - | - | 1 | 0.01 | 37 |
| 38 | - | - | - | - | - | - | - | - | 1 | - | 1 | 0.01 | 38 |
| 47 | - | - | - | 1 | - | - | - | - | - | - | 1 | 0.01 | 47 |
| 50 | - | - | - | - | - | - | 1 | - | - | - | 1 | 0.01 | 50 |
| **Grand Total** | **330** | **477** | **487** | **582** | **768** | **862** | **1026** | **1124** | **1332** | **494** | **7482** | **100.00** | **41264** |
| **%** | **4.41** | **6.38** | **6.51** | **7.78** | **10.26** | **11.52** | **13.71** | **15.02** | **17.8** | **6.60** | **100** | **AAPP*** | **5.52** |

## 5.1. AAPP-Average Author per Paper

Table 2 illustrates the year wise distribution of authorship pattern of global Brain Concussion. This study totally published 8486 papers and the authorship pattern results a total of 41264 authors. Single author contributions are accounted to 5.36 during the study period. The highest percentage of 14.33 is recorded by four authors followed by five and three authors showing 13.95 and 13.85 percentage respectively. The number of authors engaging collaborative research is found increasing year 2008 to 2017 ranging from 330 to 7482. It can be noticed that 5.52 percentages of authors collectively contribute one paper in the field of Brain Concussion.

## 5.2. Collaboration Index (CI)

Lawani proposed the Collaborative Index in 1980. It can be calculated easily, but it cannot be interpreted as a degree because it has no upper value limit. It is denoted by the formula:

$$CI = \frac{Total\ No\ of\ Authors}{Total\ No\ of\ Papers}$$

## 5.3. Degree of Collaboration

Subramanyam propounded the Degree of Collaboration, according to Subramanyam (1983)[10], a measure to figure the extent of single and multi-author papers and to interpret it as a degree.

$$DC = \frac{NM}{NS+NS}$$

$$= \frac{No.of\ Multi\ authored\ Paper}{No.of\ Single + No.of\ Multi\ authored\ Papers}$$

## 5.4. Co-authorship Index (CAI)

CAI suggested by Garg and Padhi (2001)[11] was used.

CAI is computer as follows

CAI = $\{Nij/Nio/Noi/Noo\} \times 100$

Where Nij: number of papers having j authors in year i

Nio : total output of year i

Noj : Number of papers having j authors for all years

Noo : total number of papers for all authors and all years

J = 2, (3 or 4), > = 5.

### 5.5. Collaboration Co-efficient (CC)

Ajiferuke (1988)[12] prescribed a solitary measure to gauge cooperative research and named it as collective coefficient. The accompanying formula denotes CC.

$$CC = 1 - \frac{\sum_{j}^{k}\binom{1}{j}f_j}{N}$$

### 5.6. Modified Collaboration Co-efficient (MCC)

Savanur and Srikanth (2011)[13] modified the CC and derived the MCC as follows;

$$MCC = \frac{A}{A-1} \quad 1 - \frac{\sum_{j}^{k}\binom{1}{j}f_j}{N}$$

Table 3: Analysis of collaboration factors in Brain Concussion Publications at Global Level

| Authorship pattern | 2008 | 2009 | 2010 | 2011 | 2012 | 2013 | 2014 | 2015 | 2016 | 2017 | Total |
|---|---|---|---|---|---|---|---|---|---|---|---|
| 1 | 29 | 33 | 38 | 49 | 37 | 56 | 41 | 55 | 56 | 7 | 401 |
| 2 | 45 | 72 | 65 | 75 | 81 | 89 | 107 | 125 | 153 | 32 | 844 |
| 3 | 62 | 90 | 63 | 90 | 121 | 114 | 148 | 145 | 157 | 46 | 1036 |
| 4 | 51 | 72 | 76 | 87 | 120 | 125 | 147 | 144 | 199 | 51 | 1072 |
| 5 | 38 | 74 | 69 | 61 | 119 | 130 | 156 | 165 | 171 | 61 | 1044 |
| 6 | 39 | 46 | 65 | 71 | 92 | 105 | 120 | 136 | 158 | 60 | 892 |
| 7 | 19 | 41 | 34 | 43 | 60 | 70 | 79 | 95 | 127 | 57 | 625 |
| 8 | 18 | 23 | 35 | 39 | 47 | 52 | 61 | 83 | 64 | 37 | 459 |
| 9 | 11 | 8 | 16 | 29 | 30 | 32 | 59 | 49 | 68 | 42 | 344 |
| 10 | 18 | 18 | 26 | 38 | 61 | 89 | 108 | 127 | 179 | 101 | 765 |
| Total | 330 | 477 | 487 | 582 | 768 | 862 | 1026 | 1124 | 1332 | 494 | 7482 |
| Total Author | 401 | 1688 | 3108 | 4288 | 5220 | 5352 | 4375 | 3672 | 3096 | 10064 | 41264 |
| CI | 0.82 | 0.28 | 0.16 | 0.14 | 0.15 | 0.16 | 0.23 | 0.31 | 0.43 | 0.05 | 0.18 |
| DC | 0.91 | 0.93 | 0.92 | 0.92 | 0.95 | 0.94 | 0.96 | 0.95 | 0.96 | 0.99 | 0.95 |
| CAI | 96.38 | 98.35 | 97.42 | 96.77 | 100.57 | 98.80 | 101.44 | 100.49 | 101.22 | 104.17 | 100.00 |
| CC | 0.6758 | 0.6816 | 0.6083 | 0.7085 | 0.7255 | 0.7190 | 0.7387 | 0.7335 | 0.7397 | 0.7959 | 0.7256 |
| MCC | 0.3252 | 0.3191 | 0.3925 | 0.2920 | 0.2748 | 0.2812 | 0.2615 | 0.2667 | 0.2605 | 0.2045 | 0.2744 |
| MCC-CC | 0.3506 | 0.3625 | 0.2158 | 0.4165 | 0.4507 | 0.4378 | 0.4772 | 0.4668 | 0.4792 | 0.5914 | 0.4512 |

CI-Collaborative Index, DC-Degree of Collaboration, CAI-Co-authorship Index, CC-Collaborative Co-efficient, MCC-Modified Collaborative Co-efficient

Table 3 elucidated diverse joint effort factors for the time of ten years (2008-2017). The analysis of the table incorporates CI, DC, CAI, CC and MCC. The table shows Collaborative Index at the highest in the year 2008 and lowest range at the year 2017. Mean CI during the period of study is 0.18. Subramanyam propounded the Degrees of Collaboration a measure to calculate the proportion of single and multi-author papers and to interpret it as a degree. It is found that DC was lowest at 0.91 in 2008 and highest at 0.99 in 2017. In the all the year multi-author papers are increasing, therefore the Degree of Collaboration the research period shows 0.95.

The estimation of CAI in the primary year begins with 96.38 and it increments in regard of other continuing years as multi and super author papers increment. The year 2008 onwards the values of CAI increases from 96.38 to 104.17 showing the mean of 100.00 suggesting the trend in the later years is marked with larger team sizes. In this study, CC is also lowest in 2010 showing 0.6083. It is at the highest rate of 0.7959 in 2017. The mean CC is 0.7256.

The study found MCC was lowest in 2017 when it was 0.2045. It was at the maximum value of 0.3925 in 2017. The mean MCC during the period of study was 0.2744. It is also observed from the table that the mean difference between CC and MCC is 0.4512. Least difference between CC and MCC, i.e. 0.2158 is observed the year 2010. The highest difference CC and MCC, which is0.5914, is observed in the years 2008 and 2017. It tends to be inferred that no noteworthy distinction can be seen between CC esteems, and furthermore this variety limits when the quantity of authorships increments.

**5.7. Lotka's Law**

Table 4:Lotka's law

| X | Y | X=Logx | Y=Logy | XY | $X^2$ |
|---|---|---|---|---|---|
| 1 | 16658 | 0.000000 | 4.22162 | 0.000000 | 0.000000 |
| 2 | 3397 | 0.301030 | 3.53110 | 1.062966 | 0.090619 |
| 3 | 1350 | 0.477121 | 3.13033 | 1.493548 | 0.227645 |
| 4 | 732 | 0.602060 | 2.86451 | 1.724607 | 0.362476 |
| 5 | 413 | 0.698970 | 2.61595 | 1.828471 | 0.488559 |
| 6 | 264 | 0.778151 | 2.42160 | 1.884374 | 0.605519 |
| 7 | 172 | 0.845098 | 2.23553 | 1.889240 | 0.714191 |
| 8 | 162 | 0.903090 | 2.20952 | 1.995391 | 0.815572 |
| 9 | 113 | 0.954243 | 2.12385 | 2.026671 | 0.910579 |
| 10 | 506 | 1.000000 | 2.70415 | 2.704151 | 1.000000 |
|   |   | ∑X6.559763 | 28.058163 | 16.609491 | 5.215159 |

$$n = \frac{N \sum XY - \sum X \sum Y}{N \sum X^2 - (\sum X)^2}$$

$$= \frac{10(16.609419) - (6.559763)(28.058163)}{10(5.215159) - (6.559763)^2}$$

$$= \frac{17.960709}{9.121099}$$

$$= 1.96913$$

The one of the law of Bibliometrics is Lotka's Law, which manages the recurrence of distribution by authors in some random field. The summed up type of Lotka's Law can be communicated as

Y = (C)

Where y is the quantity of authors with x articles, the type n and consistent C are parameters to be assessed from a given arrangement of author efficiency information.

While theoretical Lotka's worth is a = 2.000.

Theoretical value of 'n' 1.96913 is matched with the table value of R.Rosseau for getting C.S value -0.5974.

D-Max Value of Present Study = 0.1034

D-Max Value of Lotks's Study = 0.1314

To test the goodness of fit, weather the observed author productivity distribution is not significantly different from theoretical distribution. K-S test was applied to the data. As per the test, the greatest deviation is watched and evaluated esteem DMax is determined as follows:

$D_{max} = F(x) - E_n(x)$

a = 1.96913

Theoretical Value of C = 0.5974

Fe+ = 0.5974 (1/×1.96913)

D-Max = 0.1034

Critical Value at .0.1 level of significance

$\quad = 1.96913/\sqrt{23767}$

$\quad = 0.0128$

The theoretical values of C as 0.5974 for a =1.96913 is taken from table No. IV.6.6. in the book "Introduction to Informetrics" (Amsaveni and Batcha 2009)[14]. The K.S test is applied for the fitness of Lotka's law fits to the global Brain Concussion research output. Result indicates that the value of D – max, 0.1034 determined with Lotka's exponent, a =1.96913 for Brain Concussion which is not close and shows high to the D-max value 0.156 determined with the Lotka's type a=1 than the basic worth chose at the 0.01 degree of criticalness, 0.0128. Along these lines, distribution recurrence of the origin pursues the precise Lotka's Inverse law with the example a=1. The modified form of the inverse square law, â and C parameters as 1.96913 and 0.5974 for brain Concussion is applicable and appears to provide a good for fit.

Table 5: K-S Test

| X | Yx | Observed =Yx/∑YX | Value = ∑(YX/∑YX) | Expected Frequency | Value of Frequency/Cumulative | Difference (D) | Expected Frequency | Value of Frequency/ Cumulative | Diff |
|---|---|---|---|---|---|---|---|---|---|
| 1 | 16658 | 0.7008 | 0.7008 | 0.5974 | 0.5974 | 0.1034 | 0.6079 | 0.6079 | 0.0929 |
| 2 | 3397 | 0.1429 | 0.8437 | 0.1526 | 0.75 | 0.0097 | 0.1520 | 0.7599 | 0.0091 |
| 3 | 1350 | 0.0568 | 0.9005 | 0.0687 | 8187 | 0.0119 | 0.0675 | 0.8274 | 0.0107 |
| 4 | 732 | 0.0308 | 0.9313 | 0.0390 | 0.8577 | 0.0082 | 0.0380 | 0.8654 | 0.0072 |
| 5 | 413 | 0.0174 | 0.9487 | 0.0251 | 0.8828 | 0.0077 | 0.0243 | 0.8897 | 0.0069 |
| 6 | 264 | 0.0111 | 0.9598 | 0.0175 | 0.9003 | 0.0064 | 0.0169 | 0.9066 | 0.0058 |
| 7 | 172 | 0.0072 | 0.967 | 0.0129 | 0.9132 | 0.0057 | 0.0124 | 0.9190 | 0.0052 |
| 8 | 162 | 0.0068 | 0.9738 | 9.9532 | 10.8664 | 9.9464 | 0.0095 | 0.9285 | 0.0027 |
| 9 | 113 | 0.0047 | 0.9785 | 7.8929 | 18.7593 | 7.8882 | 0.0075 | 0.9360 | 0.0028 |
| 10 | 506 | 0.0213 | 0.9998 | 6.4141 | 25.1734 | 6.3928 | 0.0061 | 0.9421 | 0.0152 |
| Total | 23767 | | | Present study's | | D.Max =0.1034 | Lotka's | D.Max =0.0929 | |

## 5.8. Relative Growth Rate (RGR)

Relative Growth Rate means the increase in the number of articles per unit of time.

$Rt(P) = [logP(t) - logP(0)]$

## 5.9. Doubling Time

Doubling Time is defined as the time required for the articles to become double of the existing amount. It has been calculated using following formula;

Dt is given by $(t) = \dfrac{0693}{R}$

Table 6: Relative growth rate and doubling time of Brain Concussion

| Year | Output | Cum. Output | W1 | W2 | RT(p) | Mean RP(p) | Dt(p) | Mean Dt(p) |
|---|---|---|---|---|---|---|---|---|
| 2008 | 331 | - | 5.802 | - | - | | - | |
| 2009 | 477 | 808 | 6.168 | 6.695 | 0.527 | | 1.315 | |
| 2010 | 487 | 1295 | 6.188 | 7.166 | 0.978 | 0.978 | 0.709 | 0.794 |
| 2011 | 583 | 1878 | 6.368 | 7.538 | 1.17 | | 0.592 | |
| 2012 | 769 | 2647 | 6.645 | 7.881 | 1.236 | | 0.561 | |
| 2013 | 862 | 3509 | 6.759 | 8.163 | 1.404 | | 0.494 | |
| 2014 | 1026 | 4535 | 6.933 | 8.419 | 1.486 | 1.580 | 0.466 | 0.441 |
| 2015 | 1125 | 5660 | 7.026 | 8.641 | 1.615 | | 0.429 | |
| 2016 | 1332 | 6992 | 7.194 | 8.853 | 1.659 | | 0.418 | |
| 2017 | 1494 | 8486 | 7.309 | 9.046 | 1.737 | | 0.399 | |
| Total | 8486 | | | | | **1.278** | | **0.6175** |

Table 6 clearly indicates the average Relative Growth Rate and Doubling Time of articles in Brain Concussion research during the study period. It is observed that the value of relative growth rate of publications has gradually increased from 2008 (0.527) to 2017 (1.737). The doubling time of the publications gradually decreased from 1.315 (2008) to 0.399 (2017). This table can be concluded from the above analysis that relative growth Rate of articles has been gradually increased and on the other hand, doubling time of the articles has been gradually decreasing.

## 6. Conclusion

The study quantitatively identified the research productivity in the area of Brain concussion at global level over the study period of 2008-2017. The study identified the trends and characteristics of growth and collaboration pattern of Brain Concussion research output. Average growth Rate of Brain Concussion increases at the rate of 0.850. Collaborative index is noted to be the highest range at the current year 2017. Mean Collaborative Index during the period is 0.18. Lotka's Law of authorship productivity is good for application of Brain Concussion. Inverse power Law with â and C parameters as 1.96913 and 0.5974 for Brain Concussion is applicable and appears to provide a good fit. The research uncovers the way that the Brain Concussion study is one of the creating in the space of Medical Science.